\documentstyle[12pt]{article}
\begin{document}
\newcommand{\bl}{\begin{flushleft}}
\newcommand{\br}{\begin{flushright}}
\newcommand{\el}{\end{flushleft}}
\newcommand{\er}{\end{flushright}}
\newcommand{\bc}{\begin{center}}
\newcommand{\ec}{\end{center}}
\newcommand{\del}{\partial}
\newcommand{\ba}{\begin{array}}
\newcommand{\ea}{\end{array}}
\newcommand{\eq}[2]{\begin{equation}#1\label{#2}\end{equation}}
\newcommand{\eqn}[2]{\begin{eqnarray}\label{#1}#2\end{eqnarray}}
\newcommand{\nn}{\nonumber\\}
\newcommand{\grad}{\nabla}
\newcommand{\vev}[1]{\langle #1\rangle}
\newcommand{\jour}[4]{\ \em{#1}\ \bf{#2}\ \rm{(#4)\ #3}}
\newcommand{\book}[3]{\ \em{#1},\ \rm{#2\ (#3)}}
\newcommand{\listrefs}[1]{\begin{thebibliography}{999}#1
                           \end{thebibliography}}
\newcommand{\prepno}[2]{\hfill{\vbox{\hbox{#1}\hbox{#2}}}}
\newcommand{\Title}[1]{\begin{center}\LARGE{\bf #1}\end{center}
                                      \vspace{.3cm}}
\newcommand{\Author}[3]{\begin{center}\large{\bf\sf #1}\footnote{
              electronic address: #2}\\
  \smallskip\normalsize{\em #3}\\ \medskip\today\end{center}
                                   \vspace{.3cm}}
\newcommand{\Authors}[5]{\begin{center}\Large{\bf\sf #1}\footnote{
              electronic address: #3} and 
             {\bf\sf #2}\footnote{electronic address: #4}\\
   \smallskip\normalsize{\em #5}\\ \medskip\today\end{center}
                                    \vspace{.2cm}}
\newcommand{\Abstract}[1]{\begin{center}\Large{Abstract}\\
     \medskip\begin{quote}\small{#1}\end{quote}\end{center}}
\newcommand{\newsec}[1]{\section{#1}\setcounter{equation}{0}}
%
%
\newcommand{\half}{{1\over 2}}
\newcommand{\laplace}{{\kern1pt\vbox{\hrule height 1.2pt\hbox
            {\vrule width 1.2pt\hskip 3pt\vbox{\vskip 6pt}\hskip 3pt\vrule 
        width 0.6pt}\hrule height 0.6pt}\kern1pt}}
\renewcommand{\slash}[1]{{\rlap{$#1$}\thinspace /}}
\def\G{\nabla}
\def\p{\partial}
\def\D{\Delta}
\def\O{\Omega}
\def\P{\Phi}
\def\CM{{\cal M}}
\def\CL{{\cal L}}
%
\newcommand{\PR}{Phys. Rev.}
\newcommand{\PRL}{Phys. Rev. Lett.}
\newcommand{\RMP}{Rev. Mod. Phys.}
\newcommand{\NP}{Nucl. Phys.}
\newcommand{\PL}{Phys. Lett.}
\newcommand{\PREP}{Phys. Rep.}
\newcommand{\CMP}{Comm. Math. Phys.}
\newcommand{\JMP}{J. Math. Phys.}
\newcommand{\CGG}{Class. Quant. Grav.}
\newcommand{\MPL}{Mod. Phys. Lett.}
\newcommand{\IJMP}{Int. J. Mod. Phys.}
\newcommand{\AP}{Ann. Phys.}
%
%
%
%
\begin{titlepage}
\prepno{}{{\sf hep-th/9602057}}
\Title{Entropy of Extremal Dyonic Black Holes}
\Authors{A. Ghosh}{P. Mitra}{{\sf amit@tnp.saha.ernet.in}}
                            {{\sf mitra@tnp.saha.ernet.in}}
            {Saha Institute of Nuclear Physics,\\ 
             1/AF Bidhannagar, Calcutta 700 064, INDIA}
\Abstract{
For extremal charged black holes, the thermodynamic entropy is  proportional
not to the area but to the mass or charges. This is demonstrated here 
for dyonic extremal black hole solutions
of string theory. It is pointed out that these solutions have zero classical
action although the area is nonzero.
By combining the general form of the entropy allowed by
thermodynamics with recent observations in the literature it is possible
to fix the entropy almost completely.
         }
%
\end{titlepage}
%
%
%
{\bf 1.} Black hole thermodynamics has been an intriguing subject for many 
years. The
laws of classical black hole physics suggested definitions of temperature and
entropy purely by analogy with the laws of thermodynamics, but the scale of
these quantities could not be determined that way \cite{Bek}. It was only 
with the
introduction of quantum theoretical, or more precisely semiclassical, ideas
that the scale could be set in terms of Planck's constant \cite{sh}. The
temperature defined in this way, related to surface gravity, was 
later rederived in a euclidean approach where there is a requirement
of periodicity on the euclidean time coordinate if conical singularities are
to be avoided.

Apart from the obvious question about the origin of a nonzero entropy in this 
context, the expression for the entropy has itself been a cause for wonder. For 
ordinary, or what are now called non-extremal black holes, the entropy is 
proportional to the area of the horizon. Explanations have been sought to be 
given for this dependence. For instance
\cite{thooft} the statistical
entropy of matter outside the black hole is proportional to the
area of the horizon. 

There has been a lot of interest lately in the special case of extremal black
holes \cite{GMone, HHR, GMtwo}. 
The temperature and the entropy  behave differently from the
case of nonextremal black holes. Thus, when the temperature defined through 
the surface gravity is zero or infinity, it is found that there is no 
conical singularity, so that the temperature may really be arbitrary. 
Again, the thermodynamical entropy fails to be proportional to the 
area of the horizon.

Another direction which recent research has taken involves black hole
solutions of string theory. It has been possible to identify the states 
corresponding to extremal black hole solutions of string theory
\cite{duff, ashoke, strom}. This presents an
opportunity of reaching a better understanding of the  entropy of
black holes in terms  of the underlying string theory.
The entropy has indeed been calculated \cite{ashoke, strom} from the number  of
states. The result is sometimes consistent with the area formula but
sometimes nonzero even when the area of the horizon vanishes.
In the latter case, a new  interpretation of the word ``horizon'' can be
developed to match the area with the nonzero entropy. 
However, as mentioned earlier, the thermodynamic entropy of 
extremal black holes is in general
not proportional to the area. Instead of seeking an area interpretation,
the string result can be shown \cite{GMfour} to be consistent
with a modified thermodynamical formula that can be justified for 
extremal black holes. To be precise,
the expression proportional to the
mass that we advocated earlier \cite{GMtwo} (see also \cite{puri})
for the thermodynamic entropy
fits very well. While this was demonstrated explicitly for the 
electrically charged four-dimensional black
holes of \cite{ashoke}, it is clear that it also holds for the 
higher dimensional black holes of \cite{peet}.

The extremal black hole solutions considered in \cite{ashoke,peet}
are electrically charged but magnetically uncharged. Dyonic black hole
solutions were constructed in \cite{cvetic,panda} and considered 
from the point of view of state-counting in  \cite{wilczek}, where an
expression for the density of string states was proposed. For large
values of the charges, this
expression is consistent with an entropy equal to a quarter of the
area, which is of course the standard formula for non-extremal black holes.
However, the black holes under consideration are extremal, which are known to
have different euclidean topologies from ordinary black holes in general,
and the experience of \cite{GMtwo, ashoke, GMfour} indicates
the entropy to be proportional to the mass rather than the area which goes
like the square of the mass. While the calculation of \cite{strom}
suggests that the area formula may continue to hold for
some extremal black holes with constant dilaton, there is no evidence
that the formula may hold
for black holes with nontrivial dilaton fields \cite{kleb}.
In this situation,  since the dyonic black holes under consideration
do have nontrivial dilaton fields, it is  reasonable
to confront the arguments of \cite{wilczek} with other
methods which are better understood. Indeed, the clustering
inequality used in \cite{wilczek} to motivate the proposed expression for
the density of states can be easily seen to 
fail for the known density of states for
purely electrically charged black holes. The present
investigation therefore seeks to develop a formula for the 
entropy or the density of 
states by {\it avoiding the clustering argument} and using alternative 
possibilities about {\it extremal} black holes instead.

{\bf 2.} In four dimensions the massless bosonic fields of heterotic string 
obtained 
by toroidal compactification lead to an effective action with an unbroken 
$U(1)^{28}$ gauge symmetry \cite{ashoke}:
\eqn{one}{S&=&{1\over 16\pi}\int d^4x\sqrt{-g}[R-\half\p^{\mu}
\P\p_{\mu}\P+{1\over 8}Tr(\p^{\mu}\CM\CL\p_{\mu}\CM\CL)\nn
&&-{1\over 4}e^{-\P}F_{( a)\mu\nu}(\CL\CM\CL)_{ab}F^{( b)\mu\nu}
-{1\over 12}e^{-2\P}H_{\mu\nu\rho}H^{\mu\nu
\rho}].}
Here,
\eqn{two}{\CL=\left(\matrix{&I_6& \cr I_6 &&\cr &&-I_{16}\cr}\right),}
with $I$ representing an identity matrix,
$\CM$ is a symmetric 28 dimensional matrix of scalar fields satisfying
\eqn{twoa}{\quad \CM\CL\CM=\CL,}
and there are 28 gauge field tensors 
\eqn{three}{F_{(a)\mu\nu}=\p_{\mu}A_{(a)\nu}-\p_{\mu}A_{(a)\mu},~~ a=1,...,28}
as well as a third rank tensor $H$ 
\eqn{threea}{H_{\mu\nu\rho}=\p_\mu B_{\nu\rho}+2A_{(a)\mu}\CL_{ab}F_{(b)\nu\rho}
+{\rm cyclic~permutations~of~}\mu,\nu,\rho}
corresponding to an antisymmetric tensor field $B$.
The theory possesses dyonic black hole solutions. We shall 
consider the extremal dyonic solution \cite{cvetic} considered in
\cite{wilczek}.  The asymptotic forms of the fields are as follows:
\eqn{fiv}{\vev{g_{\mu\nu}}=\eta_{\mu\nu},\ \vev{e^{-\P}}=1,\ 
 \vev{B_{\mu\nu}}=0,\ \vev{A_{(a) \mu}}=0.}
In these circumstances, 
the $SL(2,R)$ (S-duality) symmetry is broken down to $SO(2)$.
The magnetic and electric charges are 28-component vectors $\vec P$ and
$\vec Q$.
It is convenient to introduce the $O(6,22)$ (T-duality) invariant magnitudes
\eqn{sieb}{Q_{R,L}=\left[\vec Q^T(\CL\CM_\infty\CL\pm\CL)\vec Q\right]^\half}
and similarly $P_{R,L}$. The ADM mass of the black hole is given by
the T- and S- duality invariant form
\eqn{acht}{4M=\left[P_R^2+Q_R^2 +2\sqrt{P_R^2Q_R^2-
[\half\vec P^T(\CL\CM_\infty\CL+\CL)\vec Q]^2}\right]^\half.}
The black hole is extremal and Bogomol'nyi-saturated. 

A specially simple form of the charge vectors corresponds to the metric
\eqn{achtnhalb}{ds^2=-{r^2\over R^2}dt^2 +{R^2\over r^2}dr^2 +R^2d\Omega^2,}
where \cite{cvetic}
\eqn{achtusw}{R^2=[(r+P_1)(r+P_2)(r+Q_1)(r+Q_2)]^\half}
with
\eqn{achtusww}{P_{R,L}= P_1\pm P_2,}
etc.

The surface gravity (at the horizon $r=0$),
if calculated from the explicit expression (\ref{achtnhalb}) for the metric, 
is zero for nonvanishing $P_1P_2Q_1Q_2$. This
way of defining the temperature 
ceases to make sense when the value obtained is zero (or
infinite). An alternative way of defining temperature 
is through the conical
singularity that tends to arise on making the time imaginary. In cases 
like this, when
the surface gravity vanishes, there is no {\it conical} singularity,
and the temperature is {\it arbitrary} \cite{HHR}.

The action can be evaluated by plugging the solution
into (\ref{one}). It is necessary
to introduce a surface term \cite{gh}. The sum of all contributions vanishes,
as is expected for an extremal black hole \cite{GMtwo}.

{\bf 3.} For nonextremal black holes, the laws of black hole physics 
suggest that
there is an entropy proportional to the area of the horizon. When the scale
is fixed by comparing the temperature thus suggested with that given by
the semiclassical calculations of \cite{sh}, the entropy turns out to be a
quarter of the area. If one is interested in an extremal black hole, one
may be tempted to regard it as a special limiting case of a sequence
of nonextremal black holes and thus infer that the same formula should
hold for the entropy. However, it was pointed out in the context of Reissner -
Nordstrom black holes \cite{HHR} that the extremal and nonextremal cases of
the euclidean version are topologically different, so that continuity
need not hold. It was also argued that the temperature in this case
is {\it arbitrary}. Subsequently 
it was shown \cite{GMtwo,puri,GMfour} that
if the derivation of an expression for the thermodynamic entropy 
along the lines of \cite{gh} is
attempted afresh for these extremal cases, with due attention paid to the
fact that {\it the mass and charges are no longer independent as in the
usual nonextremal cases,} one obtains a result proportional to the mass
of the black hole with an undetermined scale. 
Now the arguments of \cite{gh,GMtwo,GMfour} will be adapted to the 
dyonic stringy black holes. 

The first law of thermodynamics takes the form
\eqn{dreia}{\tilde T dS=dM-\vec\P_Q\cdot d\vec Q-\vec\P_P\cdot d\vec P,}
where $\vec\P_Q$ represents the chemical potential corresponding to the charge 
$\vec Q$, etc. and the temperature has been written as $\tilde T$ to
indicate the possibility of its being different from the na\"{\i}vely
vanishing temperature. It is not difficult to derive 
expressions for the chemical potential in nonextremal
cases, but we cannot use them here for two reasons: first, extremal black holes
are not continuously connected to nonextremal black holes \cite{HHR}, and
secondly, the standard expressions are calculated by differentiating the mass
with respect to charges at constant {\it area} in the anticipation that
constant area and constant entropy are synonymous, whereas in the case of
extremal black holes this relation is not necessarily valid.

Consider a process in which the mass of the black hole and all its charges
are scaled by the same factor $(1+d\epsilon)$. The relation (\ref{acht})
will continue to be satisfied. The change in entropy is given by (\ref{dreia})
to be
\eqn{dreib}{\tilde TdS=d\epsilon(M-\vec\P_Q\cdot\vec Q-\vec \P_P\cdot\vec P).}
Now 
the grand canonical thermodynamic potential 
\eqn{dreifb}{W=M-\tilde TS-\vec\P_Q\cdot\vec Q-\vec\P_P\cdot\vec P}
is related to the partition function by
\eqn{dreifa}{\exp (-{W\over\tilde T})=Z.}
Moreover, in the leading semiclassical approximation, $Z$ can be taken to be the
exponential of the negative classical action, which vanishes, as mentioned
above.
Hence  $W$ vanishes too in this approximation \cite{GMtwo}. 
Comparing (\ref{dreib}) with (\ref{dreifb}), we find
\eqn{dreigh}{\tilde TdS=d\epsilon \tilde TS.}
Thus,
\eqn{dreih}{{dS}={Sd\epsilon},}
{\it i.e.,} $S$ is a homogeneous function of the charges of degree 1.

The entropy can be expected to depend only on
combinations of the charges which are both T- and S-duality invariant.
If it is further assumed to be independent of the moduli $\CM_\infty$ 
\cite{wilczek}, the only combinations that can be involved are given by
\eqn{dreiha}{{\cal N}=2\vec P^T\CL\vec P+ 2\vec Q^T\CL\vec Q =
P_R^2-P_L^2+Q_R^2-Q_L^2}
and 
\eqn{dreihb}{{\cal A}^2=\vec P^T\CL\vec P\cdot\vec Q^T\CL\vec Q- 
(\vec P^T\CL\vec Q)^2.} 
$\cal N$ is of degree 2 and is the generalization to the dyonic case of
the combination of the charges occurring in the
expression for the {\it square} of the entropy  of electrical
black holes. Thus the square root of $\cal N$ is a very natural guess for
the answer. The other object $\cal A$ is essentially 
the area of the horizon, but again it
is of degree 2, so a square root will have to be taken. Of course, combinations
of these two objects must also be thought of. But whenever the area enters,
quarter powers of the charges
are involved if {\it negative} powers are
absent (note that in the special configuration given above,
${\cal A}=2\sqrt{P_1P_2Q_1Q_2}$).  
So it is reasonable to suppose that only $\cal N$ 
appears  in the expression for the entropy. This fixes the expression upto
a constant. If the further guess is made that the constant is the same as
in the case of purely electrically charged black
holes, one is led to
\eqn{dreii}{S={\rm const}\times\sqrt{P_R^2-P_L^2+Q_R^2-Q_L^2}
\sim 4\pi\sqrt{P_R^2-P_L^2+Q_R^2-Q_L^2}.}
Like the expression proposed in \cite{wilczek}, this reduces to the one
found by counting string states in \cite{ashoke} when there is no magnetic
charge, but unlike their proposal, it does not go over to the area when both
electric and magnetic charges are large. We believe that something like this
should be valid for these extremal black holes, for which there is no a priori 
reason for the area formula to hold. 

For completeness, one should also calculate the entropy of a matter field in the
background of this dyonic black hole. The entropy thus obtained is called
entanglement entropy in the literature. This is entirely different from the
thermodynamic entropy that has been discussed above, but as it has been widely
considered in the context of black hole entropy, it is reasonable to compare
it with the forms suggested above and in \cite{wilczek}.
In the spirit of \cite{thooft,
GMone} we take a free scalar field and try to calculate its
free energy when the support of the fields is reduced to the region
only outside the horizon. We need to cut-off the arbitrarily high-energy
modes that appear into the partition function and this in effect gives rise
to the thermalization of the field outside. First we observe that the
metric behaves very much like the extremal Reissner-Nordstrom one and hence
the essential form of the entanglement entropy is expected to be the same.
Secondly we observe that the metric in the Euclidean $r-t$-plane near the 
horizon cannot be brought to a form of a flat cone with a finite conformal 
transformation. This means that the temperature which in this case would
correspond to the periodicity of the euclidean time coordinate remains
arbitrary. This is exactly what happens in the case of an extremal Reissner-
Nordstrom black hole. The free energy  can be seen from the general formula
given in \cite{puri} to behave exactly the same way as in \cite{GMtwo}
\eqn{}{{\cal F}={2\pi^3\over 135\beta^4}{{\cal A}\over\epsilon^3}}
where $\epsilon$ is a cut-off from the horizon. If we introduce
a proper distance then the corresponding entropy has a huge exponential
divergence and the prefactor in front of it is ${\cal A}^3$. So there is
no suggestion from this entangled point of view that the form
of the entropy is proportional to either the area or  $\sqrt{\cal N}$.
However, we do not think that this entropy has to have a connection with the
entropy of the black hole.

To sum up, we have discussed the entropy of extremal dyonic black holes in the
spirit of \cite{wilczek} but with two major differences: we have avoided using
their clustering argument but have taken the extremality into
account while adapting the procedure of \cite{gh} to this case. This results
in a proposal for the entropy different from  the one in \cite{wilczek}. We
then considered the entropy of matter in the black hole background because 
of its popularity, but it disagrees with {\it both} of the proposals. A better
understanding of the meaning of the string calculations of \cite{strom} may
throw more light on these questions, in particular on the discontinuity in the 
extremal limit.

%
\newpage
\listrefs{
\bibitem{Bek} J. Bekenstein,\jour{\PR}{D9}{3292}{1974}.
\bibitem{sh} S. W. Hawking,\jour{\CMP}{15}{1375}{1975}.
\bibitem{thooft} G. 't Hooft,\jour{\NP}{B256}{727}{1985}.
\bibitem{GMone} A. Ghosh and P. Mitra,\jour{\PRL}{73}{2521}{1994}.
\bibitem{HHR} S. Hawking, G. Horowitz and S. F. Ross,\jour{\PR}{D51}
                {4302}{1995}.
\bibitem{GMtwo} A. Ghosh and P. Mitra,\jour{\PL}{B357}{295}{1995}.
\bibitem{duff} M. Duff, R. R. Khuri, R. Minasian and J. Rahmfeld,
                  \jour{\NP}{B418}{195}{1994}; M. Duff and J. Rahmfeld,
                  \jour{\PL}{B345}{441}{1995}. 
\bibitem{ashoke} A. Sen, \jour{\MPL}{A10}{2081}{1995}.
\bibitem{strom} A. Strominger and C. Vafa, hep-th/9601029. 
\bibitem{GMfour} A. Ghosh and P. Mitra, hep-th/9509090.
\bibitem{puri} P. Mitra, gr-qc/9503042, to appear in Proceedings of the
               Puri workshop on {\em Physics at the Planck scale}, {\it ed.}
               by A. Kumar and J. Maharana, World Scientific, Singapore.
\bibitem{peet} A. Peet, hep-th/9506200.
\bibitem{cvetic} M. Cvetic and D. Youm, \jour{\PR}{D53}{584}{1996}.
\bibitem{panda} D. Jatkar, S. Mukherji and S. Panda, hep-th/9512157.
\bibitem{wilczek} F. Larsen and F. Wilczek, \jour{\PL}{B375}{37}{1996}. 
\bibitem{kleb} I. Klebanov and A. Tseytlin, hep-th/9604089.
\bibitem{gh} G. Gibbons and S. W. Hawking, \jour{\PR}{D15}{2752}{1977}.
            }
%
\end{document}